\documentclass[twocolumn,showpacs,10pt,amsmath,amssymb]{revtex4}
\usepackage{graphicx}

\usepackage{amsmath,amsfonts,amssymb,bm}

\begin{document}

\draft
\title{Random walk in nonhomogeneous environments: A possible approach to human and animal mobility}

\author
{Tomasz Srokowski}

\affiliation{
 Institute of Nuclear Physics, Polish Academy of Sciences, PL -- 31-342
Krak\'ow,
Poland }

\date{\today}

\begin{abstract}

The random walk process in a nonhomogeneous medium, characterised by a L\'evy stable distribution of jump length, is discussed. 
The width depends on a position: either before the jump or after that. In the latter case, the density slope is affected 
by the variable width and the variance may be finite; then all kinds of the anomalous diffusion are predicted. 
In the former case, only the time characteristics are sensitive to the variable width. 
%while the former case  resolves itself to a problem with a variable jumping rate, 
The corresponding Langevin equation with different interpretations of the multiplicative noise is discussed. 
The dependence of the distribution width on position after jump is interpreted in terms of cognitive abilities and  
related to such problems as migration in a human population and foraging habits of animals. 
%and results compared with predictions of the random walk. 

\end{abstract} 

\pacs{05.40.Fb,02.50.Ey}
%,89.65.-s

\maketitle

Usually one assumes that a waiting time and a jump-size distributions (JSD) in a continuous time random walk model (CTRW) are either 
independent \cite{met}, and can be separated, or coupled (L\'evy walks) \cite{zabu}. Those approaches may not be sufficient if one considers, 
in particular, a particle that moves in a disordered medium with heterogeneously distributed traps making 
the waiting time dependent on the current position. That effect can be taken into account by introducing a variable subdiffusion 
exponent \cite{che} or a variable intensity of a random time distribution \cite{sro15}. The position-dependent waiting time 
influences the time characteristics of the system. However, JSD may also be affected by the heterogeneous medium structure and depend 
on the position. Taking into account that dependence is a subject of the present paper. We assume JSD in a L\'evy stable form. 

The problems one can have in mind in this context include a mobility pattern of people and animals. 
It is well-known, and demonstrated, e.g., for spider monkeys \cite{ramos} and marine predators \cite{sims}, 
that the animal trajectory is often governed by a L\'evy stable distribution; in fact, this distribution corresponds 
to the optimal search strategy and is evolutionary optimal \cite{sims,bart}. 
Moreover, the L\'evy flights occur in many areas of science \cite{shles}. 
Studies of human movements are of special importance: they range from efforts to improve a traffic structure to preventing 
the spread of infectious diseases. The analysis of the dispersal of bank notes indicates that 
the length of human travels obeys the non-Gaussian L\'evy statistics \cite{gei} and is governed by CTRW. In contrast to that 
purely random picture, the study of trajectories of mobile phone users reveals reproducible patterns 
with many returns to the same places in their daily routine \cite{song}. 
%They reflect the travel habits of individual users in their daily routine and reflect a heterogeneous environment structure . 
The long term spatial and temporal scaling patterns are observed in that analysis, as well as systematic deviations from CTRW. 
However, JSD may depend on the local conditions: for example, if the walker is looking for a job and just now abides in a region that offers 
many workplaces, the jumps are shorter than those predicted by the unbiased distribution. On the other hand, 
scarcely populated and poor regions require longer jumps. Similarly, the movement of predators depends on geographical prey distributions 
while primates use mental maps of resource location to plan their jumps which can make them nearly deterministic \cite{sims} and characterised 
by a very slow diffusion \cite{boy1}. 
In any case, JSD, obeying, asymptotically, a scaling form, is not purely random: it depends on the environment structure. 

We assume JSD $Q(\xi)$ in a stable and symmetric form with a stability index $\alpha$ ($0<\alpha<2$) and a Fourier transform 
${\cal F}[Q(\xi)]=\hbox{e}^{-|k|^\alpha}$. 
The bias due to the local conditions enters $Q$ by a modification of the width by a positive and symmetric function $f(x)$ and JSD reads 
\begin{equation}
\label{q1}
Q_{x'}(x-x')=\frac{1}{f(x')}Q[|x-x'|/f(x')]. 
\end{equation}
It corresponds to a jump from $x'$ to $x$: $x=x'+f(x')\xi$, where the random number $\xi$ is sampled from $Q(\xi)$. 
Therefore, $1/f(x)$ has a sense of the concentration of the favoured places: the larger this concentration the smaller the jump length. 
The time elapsing between subsequent jumps is given by a waiting time distribution which we assume as 
a Poissonian with a variable rate $\nu(x)$ \cite{uwa}. 
If this time is sufficiently large the travel time may be neglected (as is usually the case, e.g., if one takes up a job). 
The evolution of the density distribution is governed by a master equation (ME) \cite{gar} 
\begin{equation}
\label{meqg}
\partial p(x,t)/\partial t=\int dx'[W(x|x')p(x',t)-W(x'|x)p(x,t)], 
\end{equation}
where $W(x|x')$ is a transition probability per unit time and, in the our case, $W(x|x')=Q_{x'}(x-x')\nu(x')$. 
Then 
\begin{eqnarray}
\label{meq}
\frac{\partial p(x,t)}{\partial t}&=&-\nu(x) p(x,t)\\
&+&\int\frac{1}{f(x')}Q\left(\frac{|x-x'|}{f(x')}\right)\nu(x')p(x',t)dx'.\nonumber
\end{eqnarray}
\begin{center}
\begin{figure}
\includegraphics[width=90mm]{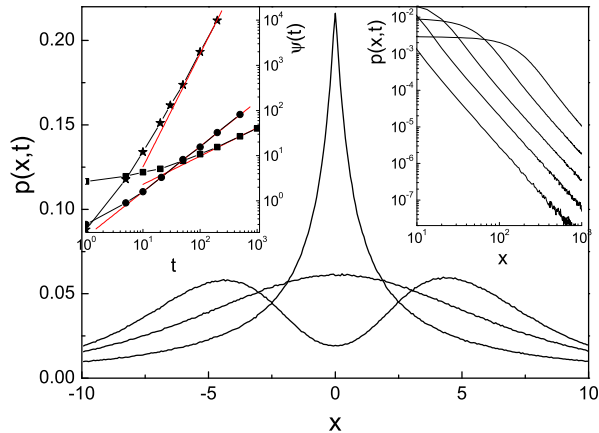}
\caption{(Colour online) Distributions corresponding to Eq.(\ref{meq}) evaluated from trajectory simulations with $\alpha=1.5$, $\eta=0$ and 
$\theta=-0.5, 0, 0.5$ (from top to bottom) at $t=10$. 
Right inset: the tail of the distributions at $t=1$ for $\theta=$-0.5, 0.5, 1, 1.5 and 2 (from left to right). The initial 
condition: $p(x,0)=\delta(x-0.1)$. Left inset: time-dependence of densities corresponding to Eq.(\ref{meq}) 
evaluated from trajectory simulations for $\alpha=1.5$ and two values of $\theta$: 0.6 (squares) and -0.6 (stars). 
Dots refer to the case of Eq.(\ref{meqai}) with $\theta=0.6$ and $\eta=0$. 
The red straight lines mark the analytical results.} 
\end{figure}
\end{center}  

Let us consider a diffusion limit of small wave numbers which correspond to large $|\xi|$ and substitute $Q(\xi)$ by 
its algebraic tails $A_\alpha |\xi|^{-1-\alpha}$ ($A_\alpha=$const). 
Then $W(x|x')=A_\alpha f^\alpha(x')|x-x'|^{-1-\alpha}\nu(x')$. The inserting of this expression 
into Eq.(\ref{meqg}) and taking into account the normalisation of $Q(\xi)$ yields a new ME, 
\begin{eqnarray}
\label{meqa}
\frac{\partial p(x,t)}{\partial t}&=&-\nu(x) f^\alpha(x)p(x,t)\nonumber\\
&+&\int \nu(x') f^\alpha(x') Q(x-x')p(x',t)dx', 
\end{eqnarray}
which approximates Eq.(\ref{meq}) and is valid for large $|x|$. 
Note that Eq.(\ref{meqa}) describes a process with an effective position-dependent jumping rate $\nu_f(x)=\nu(x) f^\alpha(x)$ \cite{sro06}: 
asymptotically, the rescaling of the jump length appears equivalent to a modification of the time characteristics. 
%That new rate follows from the scaling property: the scaling of a path $l$. 
In the diffusion limit, Eq.(\ref{meqa}) resolves itself to the Fokker-Planck equation (FPE) \cite{sro06}, 
\begin{equation}
\label{frace}
\frac{\partial p(x,t)}{\partial t}=\frac{\partial^\alpha[\nu_f(x)p(x,t)]}{\partial|x|^\alpha},
\end{equation}
where the fractional Riesz-Weyl derivative is defined by the Fourier transform, 
$\frac{\partial^\alpha}{\partial|x|^\alpha}=-{\cal F}^{-1}[|k|^\alpha]$. 
The Fourier transforming of Eq.(\ref{frace}) yields the first non-constant term as $|k|^\alpha$ which implies the asymptotics 
$p(x,t)\propto |x|^{-1-\alpha}$ for any $f(x)$ and $\nu(x)$ with finite means. To derive the dependence of the density on time, one has to 
assume a specific form of $\nu_f(x)$. We take 
\begin{equation}
\label{fodx}
f(x)=|x|^{-\theta}~~~\hbox{and}~~~\nu(x)=|x|^{-\eta} 
\end{equation}
for $|x|\gg1$. The power-law form of $\nu_f(x)$ is natural for problems with scaling. 
It is compatible with the power-law statistics observed in the migration dynamics \cite{gei,song} 
and the foraging habits of both primates \cite{boy} and marine predators \cite{sims}. More general, it corresponds to a hypothesis 
that the scaling laws describe a fundamental order in living and complex systems \cite{kello}. 
Moreover, the power-law form of the diffusion coefficient is appropriate to describe diffusion on fractals \cite{osh} 
and it would be natural in disordered systems where faults often exhibit a fractal structure and may serve as traps; 
e.g., in geology such a network of fractures is responsible for transport in a rock \cite{pai}. 
The interpretation of Eq.(\ref{fodx}) is obvious: if $\theta>0$, the concentration of favoured places rises with the distance 
and the walker proceeds with effectively smaller steps; otherwise, the probability of long jumps rises. 
The asymptotic solution of Eq.(\ref{frace}) follows from its scaling properties, 
\begin{equation}
\label{solp0}
p(x,t)\sim \psi(t)|x|^{-1-\alpha}
\end{equation}
where $\psi(t)\propto t^{1/(1+\theta+\eta/\alpha)}$. 

The density distributions inferred from the trajectory simulations are presented in Fig.1. The tails exhibit the power-law shape, Eq.(\ref{solp0}), 
the density at the origin reveals a gap for $\theta>0$ and has a cusp for $\theta<0$. The time dependence of the tails, $\psi(t)$, 
converges with time to the analytical result, Eq.(\ref{solp0}), that is also presented in the figure. 

Up to now, we have assumed that the width of JSD is determined by the current position: the walker stays longer 
in the region of high concentration of the favoured places. However, if we are dealing with the migration of humans, 
the walker is expected to be focused not on the present position but on the target. Jumping over a stream, one cares about the landing place. 
The walker reckons, makes plans and predictions about consequences of a jump, using 
available informations, and checks whether the next possible location in the neighbourhood is optimal. 
If this is the case, the walker does not need to search in the distance and the jumps become relatively shorter. 
Otherwise, the walker has to try longer jumps and the outcome may be uncertain due to a limited knowledge of the distant regions. 
Similarly, the maritime predators have an incomplete knowledge of resource location; 
if it exceeds the sensory detection range, they must initiate searches aimed at traversing larger distances \cite{sims}. 
%and lead to an uncertain outcome (in particular, if the walker is an foreigner and has a limited knowledge about possibilities in a global scale). 
JSD is flat in those cases; jumps which end up at a given position $x$ are rather accidental and originate from a large basin. 

In order to take those effects into account, we assume JSD, instead of Eq.(\ref{q1}), in the form, 
\begin{equation}
\label{q2}
Q_{x}(x-x')=\frac{1}{f(x)}Q[|x-x'|/f(x)]; 
\end{equation}
$Q_{x}(x-x')dx'$ means a probability that a jump which ends up at $x$ started within the interval $(x',x'+dx')$. 
%now, JSD is normalised in respect to $x'$ and the target position $x$ serves as a parameter: there is always a starting position 
%corresponding to the given arrival point $x$. 
ME reads, 
\begin{equation}
\label{meqai}
\begin{split}
\frac{\partial p(x,t)}{\partial t}&=\int\frac{1}{f(x)}Q\left(\frac{|x-x'|}{f(x)}\right)\nu(x')p(x',t)dx'\\
&-\nu(x)p(x,t)\int\frac{1}{f(x')}Q\left(\frac{|x-x'|}{f(x')}\right)dx'. 
\end{split}
\end{equation}
Eq.(\ref{meqai}) is complicated in general and we restrict our analysis to the asymptotic regime. 
Approximating $Q(\xi)$ by its power-law tails, the first integral yields 
\begin{equation}
\label{i1}
A_\alpha f^\alpha(x)\int|\xi|^{-1-\alpha}\nu(x-\xi)p(x-\xi,t)d\xi,
\end{equation}
and the second one, 
\begin{equation}
\label{i2}
A_\alpha \nu(x)p(x,t)\int|\xi|^{-1-\alpha}f^\alpha(x-\xi)d\xi\approx f^\alpha(x)\nu(x)p(x,t), 
\end{equation}
where we have taken into account that the contribution to the integral from large $\xi$ is negligible for large $|x|$, which 
is the case when $f(x)$ does not rise too strongly. Then we take 
the Fourier transform, expand all functions in the fractional powers of $|k|$ and preserve only the first non-constant term. 
Since we did not care about the region of small $|x|$, the procedure destroyed the normalisation. It can be restored by adjusting 
the $k$-independent term in the characteristic function. 
Since ${\cal F}[Q(x)]\sim 1-|k|^\alpha$, the Fourier transformed equation takes the form, 
\begin{equation}
\label{fracek1}
\frac{\partial}{\partial t}\widetilde p_1(k,t)=-|k|^\alpha{\cal F}[f^\alpha(x)\nu(x)p_1(x,t)]+\frac{\partial}{\partial t}\langle f^{-\alpha}\rangle_p  
\end{equation}
and 
\begin{equation}
\label{psol}
p(x,t)=f^\alpha(x) p_1(x,t). 
\end{equation}
Finally, the inversion of the transform yields, 
\begin{equation}
\label{frace1}
f^{-\alpha}(x)\frac{\partial p(x,t)}{\partial t}=
\frac{\partial^\alpha[\nu(x)p(x,t)]}{\partial|x|^\alpha}+\frac{\partial}{\partial t}\langle f^{-\alpha}\rangle\delta(x). 
\end{equation}
\begin{center}
\begin{figure}
\includegraphics[width=10cm]{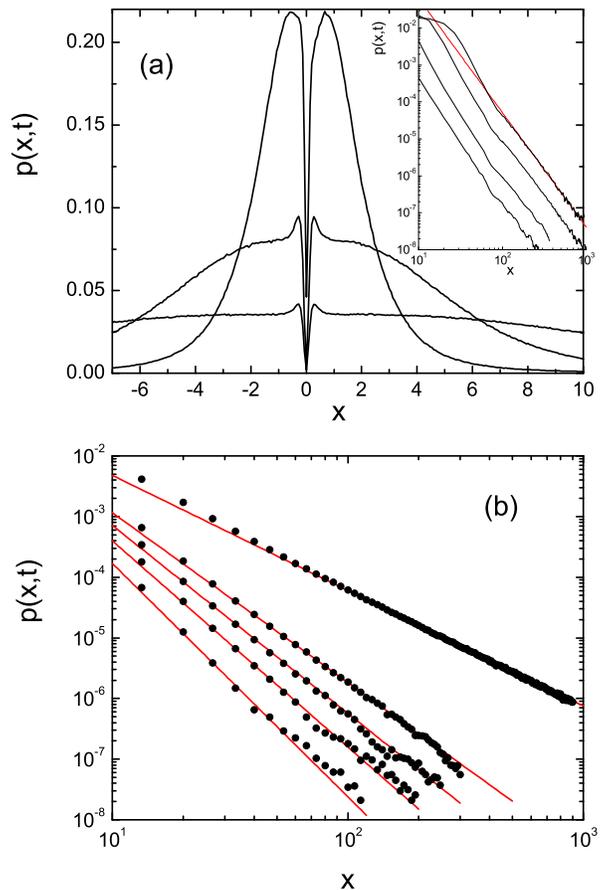}
\caption{(Colour online) (a) Time evolution of densities corresponding to Eq.(\ref{meqai}) evaluated from trajectory simulations with $\alpha=1.5$, 
$\theta=0.3$ and $\eta=0$ for $t=1$, 10, 50 (from top to bottom). 
Inset: the evolution for $\theta=0.6$ $t=$ 1, 10 100 and 500 (from left to right). 
(b) The tails for $\theta=$ 0.9, 0.6, 0.4, 0.2 and -0.4 (from left to right). The red straight lines mark the dependence 
$x^{-1-\alpha-\alpha\theta}$.} 
\end{figure}
\end{center}  

Some properties of the solution of Eq.(\ref{frace1}) can be concluded without assuming a specific form of the functions $f(x)$ and $\nu(x)$. 
Asymptotically, ${\cal F}[f^\alpha(x)\nu(x)p_1(x,t)]\sim \langle \nu(x)\rangle_p+o(k^0)$ which implies, provided the above mean 
exists, the dependence 
%$p_1(x,t)\sim\psi_1(t)|x|^{-1-\alpha}$ and 
%the tails of $p(x,t)$ are modified according to Eq.(\ref{psol}), 
\begin{equation}
\label{sol1}
p(x,t)\sim\psi(t)f^\alpha(x)|x|^{-1-\alpha}. 
\end{equation}
We observe the essential difference compared to Eq.(\ref{solp0}): $f(x)$ changes the shape 
of the tail and, if it is a decreasing function, the variance may be finite. 

The function $\psi(t)$ can be analytically derived if $f(x)$ and $\nu(x)$ obey the power-law form, namely Eq.(\ref{fodx}) with $\theta>-1$. 
Then the solutions admit the scaling form and, in particular, the influence of the medium heterogeneity 
on the diffusion properties can be easily inferred. 
The indexes $\theta$ and $\eta$ may be both positive and negative and we assume that they are mutually independent. 
The case $\theta>0$ is preferred if we expect the walker proceeds towards more favoured places. 
%but one can expect that the increasing concentration of the favoured places with $|x|$ ($\theta>0$) could be correlated with 
%a shorter dwelling time (e.g. one can find a better job in the neighbourhood), i.e. $\eta<0$. 
%Since 
%the walker usually proceeds towards more favoured places, one can expect that $\theta>0$. In an advantageous environment, it is not 
%necessary to stay long in one place (e.g. if one can find a better job in a neighbourhood); 
%therefore, the jumping rate is expected to rise with position ($\eta<0$). 
The scaling solution effectively depends on one variable, 
\begin{equation}
\label{pscal}
p_1(x,t)=a_1(t)p_1[a_1(t)x]~~~\hbox{and}~~~p(x,t)=a(t)p[a(t)x];  
\end{equation}
this implies $p_1(x,t)\sim a_1^{-\alpha}(t)|x|^{-1-\alpha}$ and $a^{-\alpha(1+\theta)}(t)=a_1^{-\alpha}(t)$. 
Since $\widetilde p_1(k,t)\sim \langle |x|^\theta\rangle_p-ca_1^{-\alpha}(t)|k|^\alpha$, where $c=$const, and 
\begin{equation}
\label{prstr}
%\begin{split}
\int|x|^{-\alpha\theta-\eta}p_1(x,t)dx=
%\int|x|^{-\eta}p(x,t)dx=\\
a^\eta\int|x|^{-\eta}p(x)dx\equiv h_0 a^\eta, 
%\end{split}
\end{equation}
we can determine $a(t)$ from Eq.(\ref{fracek1}) by means of a differential equation, 
\begin{equation}
\label{difeq}
\dot a(t)=-\frac{h_0}{\alpha(1+\theta)}a^{\alpha+\alpha\theta+\eta+1}(t). 
\end{equation}
Its solution reads, 
\begin{equation}
\label{sola}
a(t)=\psi^{-1/(\alpha+\alpha\theta)}\propto t^{-1/(\alpha+\alpha\theta+\eta)}, 
\end{equation}
which, finally, produces the solution of Eq.(\ref{frace1}), 
\begin{equation}
\label{solp}
p(x,t)\propto t^{\frac{1+\theta}{1+\theta+\eta/\alpha}}|x|^{-1-\alpha-\alpha\theta}. 
\end{equation}
Effectively, $p(x,t)$ depends on two parameters: $\alpha(1+\theta)$ and $\eta$. 
Eq.(\ref{solp}) is valid if the mean $h_0$ exists, i.e. when $\alpha(1+\theta)+\eta>0$, and the normalisation condition 
implies $\alpha(1+\theta)>0$. The slope depends on $\theta$, in contrast to Eq.(\ref{solp0}), but is independent of 
the waiting time parameter $\eta$. Eq.(\ref{solp}) represents either the asymptotics of the stable distribution with 
a stability index $\alpha(1+\theta)<2$ or a fast falling tail with slope $>3$. 

Performing the trajectory simulations for the above problem, 
%are more difficult than for the case of Eq.(\ref{q1}) since 
we deal with the stochastic equation, 
\begin{equation}
\label{steq}
x=x'+f(x)\xi,
\end{equation}
which is not explicit and we have to numerically solve a nonlinear equation at each jump. 
The density distributions resulting from such numerical analysis are presented in Fig.2. 
The slope of the tails depends on $\theta$ obeying the dependence (\ref{solp}) while at the origin the density falls to zero 
for $\theta>0$. The time dependence of the density is shown in Fig.1 (left inset). 

According to Eq.(\ref{solp}), the variance is finite if $\alpha(1+\theta)>2$ and then it determines the diffusion properties. 
Variance follows from a simple scaling: 
\begin{equation}
\label{var}
\langle x^2\rangle(t)=a(t)\int x^2 p[a(t)x]dx\propto a(t)^{-2}\propto t^\mu, 
\end{equation}
where $\mu={2/(\alpha+\alpha\theta+\eta)}$. We observe the normal diffusion ($\mu=1$) and both kinds of the anomalous behaviour: 
a subdiffusion ($\mu<1$) and an enhanced diffusion ($\mu>1$). 
In particular, the normal diffusion emerges if the decline of the function $f(x)$ 
is exactly compensated by the rising jumping rate, $-\eta=\alpha(1+\theta)-2$. 
However, this does not mean the Gaussian statistics: the distribution looks like that in Fig.2. 
The variance for all three diffusion regimes is compared with the numerical simulations in Fig.3. 
\begin{center}
\begin{figure}
\includegraphics[width=95mm]{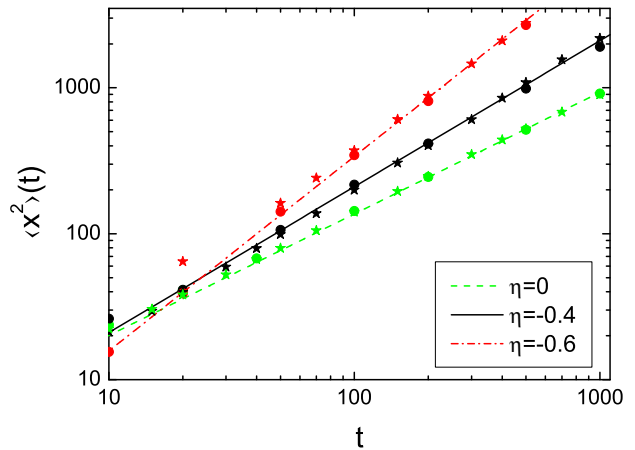}
\caption{(Colour online) Variance as a function of time evaluated from CTRW trajectory simulations for $\alpha=1.5$ and 
$\theta=0.6$ (points). The slope of straight lines obeys the dependence (\ref{var}) with $\mu=1.33$, 1 and 0.833. Stars mark rescaled results 
of the numerical solving of Eq.(\ref{lan}) where $\nu(x)$ was evaluated before the random kick and $f(x)$ after that (anti-It\^o interpretation).} 
\end{figure}
\end{center}  

In the continuous limit, Eq.(\ref{steq}) becomes a Langevin equation with a multiplicative noise, 
\begin{equation}
\label{lan}
dx(t)=f(x)\nu^{1/\alpha}(x)\xi(dt),
\end{equation}
where $\nu(x)$ is evaluated just {\it before} the random kick (the It\^o interpretation) \cite{sro15} 
and $f(x)$ just {\it after} that (the anti-It\^o interpretation); $\xi$ stands for a white noise. 
Eq.(\ref{lan}) has been numerically 
solved and the time-dependence of the variance is presented in Fig.3: the slope agrees with Eq.(\ref{var}). 
On the other hand, $\xi$ may be regarded as a limit of a correlated noise \cite{sro12} which implies another approach to 
the multiplicative noise with respect to $f(x)$. 
Then we may change the variable in a standard way, $x\rightarrow y=\int_0^x dx'/f(x')$, and Eq.(\ref{lan}) becomes 
$dy(t)=\nu^{1/\alpha}[x(y)]\xi(dt)$ which resolves itself to FPE, 
\begin{equation}
\label{fracel}
\frac{\partial p(y,t)}{\partial t}=\frac{\partial^\alpha[\nu(x(y))p(y,t)]}{\partial|y|^\alpha}.
\end{equation}
The asymptotic solution of Eq.(\ref{fracel}), expressed by the original variable $x$, agrees with the solution of ME (\ref{meqai}), Eq.(\ref{sol1}), 
for any $f(x)$ and $\nu(x)$ with finite means. Therefore, the application of ordinary rules of the calculus to the Langevin equation 
results in the same shape of the tails as in the anti-It\^o interpretation. 
Moreover, if $f(x)$ and $\nu(x)$ assume the algebraic forms, Eq.(\ref{fodx}), 
%$y(x)=|x|^{1+\theta}\hbox{sgn}x/(1+\theta)$ 
$\nu(y)=[(1+\theta)|y|]^{-\eta/(1+\theta)}$ and we obtain Eq.(\ref{solp}) while the variance obeys Eq.(\ref{var}) \cite{sro15}. 
In the Gaussian case, the above procedure of the variable change corresponds to the Stratonovich interpretation when the multiplicative term 
is evaluated in the middle point. The numerical analysis \cite{sro09} suggests that this correspondence may also hold 
for the L\'evy flights but, in general, the multiplication noise interpretation for the L\'evy flights is still an open problem. 
The Langevin equation in the Stratonovich interpretation has been applied to study a Lotka-Volterra system of two competing species \cite{cogn}. 
Moreover, the multiplicative L\'evy noise can be considered in terms of a Marcus interpretation \cite{marcus}. 
We conclude that agreement of the solutions of the Langevin equation 
with those of ME (\ref{meqai}) seems robust with respect to a particular interpretation. However, 
it must be different from It\^o for which interpretation, in turn, 
%means that $f(x)$ is evaluated just before the force acts and 
Eq.(\ref{lan}) leads to FPE in the form (\ref{frace}) with the effective diffusion coefficient $\nu_f(x)=\nu(x)f^\alpha(x)$ \cite{sche}; 
this case corresponds to ME (\ref{meq}). In the asymptotic limit, solution of FPE agrees with that of ME and 
is given by Eq.(\ref{solp0}) \cite{sro06,sro09}. 
%slope of the tail appears independent of $\theta$ and $\eta$, variance is infinite. 

In summary, we have discussed the random walk process with the position-dependent, $\alpha$-stable JSD $Q_x$ which reflects 
a heterogeneous medium structure. There are two possibilities: (1) $Q_x$ depends on the position {\it before} the jump, 
then the problem asymptotically resolves itself to the ordinary CTRW but with a variable effective jumping rate; 
(2) $Q_x$ depends on the position {\it after} the jump. Now, not only the time characteristics but 
also the asymptotic shape of the distribution is affected by the $x$-dependent width of $Q_x$. 
Those two cases lead to qualitatively different predictions: while for (1) we observe the L\'evy stable asymptotics 
with the stability index $\alpha$ and infinite variance (which often is problematic for the physical reasons), 
for (2) the slope of the tail may be large and the variance finite without any 
truncation of the distribution; we observe all kinds of the anomalous diffusion. The agreement of the Langevin equation solution 
with that for the random walk indicates that the application of the ordinary rules of the calculus corresponds to the anti-It\^o 
interpretation of the multiplicative noise: both formalisms yield the same form of the tails. 
The case (2) is natural when we consider movements of humans and animals which 
entail cognitive abilities: jumps are planed {\it a priori} on the basis of knowledge, intuition and outcome predictions.


\begin{thebibliography}{99}

\bibitem{met}
R. Metzler and J. Klafter, Phys. Rep. {\bf 339}, 1 (2000). 

\bibitem{zabu}
V. Zaburdaev, S. Denisov, and J. Klafter, Rev. Mod. Phys. {\bf 87}, 483 (2015). 

\bibitem{che} 
A. V. Chechkin, R. Gorenflo, and I. M. Sokolov, J. Phys. A: Math. Gen. {\bf 38}, L679 (2005); 
B. A. Stickler and E. Schachinger, Phys. Rev. E {\bf 84}, 021116 (2011).  

\bibitem{sro15}
T. Srokowski, Phys. Rev. E {\bf 89}, 030102(R) (2014); {\it ibid} {\bf 91}, 052141 (2015). 

\bibitem{ramos}
G. Ramos-Fern\'andez, J. L. Mateos, O. Miramontes, H. Larralde, G. Cocho, and B. Ayala-Orozco, Behav. Ecol. Sociobiol. {\bf 55}, 223 (2004). 

\bibitem{sims}
D. W. Sims {\it et al.}, Nature {\bf 451}, 1098 (2008). 

\bibitem{bart}
F. Bartumeus, M. G. E. da Luz, G. M. Viswanathan, end J. Catalan, Ecology {\bf 86}, 3078 (2005).

\bibitem{shles}
{\it L\'evy flights and related topics in physics}, edited by M. F. Shlesinger, G. M. Zaslavsky, and J. Frisch (Springer Verlag, Berlin, 1995); 
{\it L\'evy processes: Theory and applications}, edited by O. E. Barndorff-Nielsen, T. Mikosch, and S. I. Resnick (Birkh\"auser, Boston, 2001).  

\bibitem{gei}
D. Brockmann, L. Hufnagel, and T. Geisel, Nature {\bf 439}, 462 (2006). 

\bibitem{song}
C. Song, T. Koren, P. Wang, and A.-L. Barab'asi, Nat. Phys. {\bf 6}, 818 (2010). 

\bibitem{boy1}
D. Boyer and C. Solis-Salas, Phys. Rev. Lett. {\bf 112}, 240601 (2014). 

\bibitem{uwa}
If the waiting time distribution has a heavy tail given by the index $\beta<1$, the density follows from the solution of 
the Poissonian case by a rescaling of the Laplace transform: $p(x,s)\rightarrow s^{\beta-1}p(x,s^\beta)$. 

\bibitem{gar}
C. W. Gardiner, {\it Handbook of Stochastic Methods for Physics, Chemistry
and the Natural Sciences} (Springer-Verlag, Berlin, 1985).

\bibitem{sro06}
T. Srokowski and A. Kami\'nska, Phys. Rev. E {\bf 74}, 021103 (2006). 

\bibitem{boy}
D. Boyer {\it et al.}, Proc. R. Soc. Lond. B {\bf 273}, 1743 (2006). 

\bibitem{kello}
C. T. Kello {\it et al.}, Trends Cogn. Sci. {\bf 14}, 223 (2010). 

\bibitem{osh}
B. O'Shaughnessy and I. Procaccia, Phys. Rev. Lett. {\bf 54}, 455 (1985). 

\bibitem{pai}
S. Painter, V. Cvetkovic, and J.-O. Selroos, Phys. Rev. E {\bf 57}, 6917 (1998); 
T. A. Tafti, M. Sahimi, F. Aminzadeh, and C. G. Sammis, Phys. Rev. E {\bf 87}, 032152 (2013). 

\bibitem{sro12}
T. Srokowski, Phys. Rev. E {\bf 85}, 021118 (2012). 

\bibitem{sro09}
T. Srokowski, Phys. Rev. E {\bf 80}, 051113 (2009); {\it ibid} {\bf 81}, 051110 (2010). 

\bibitem{cogn}
A. La Cognata, D. Valenti, A. A. Dubkov, and B. Spagnolo, 
Phys. Rev. E {\bf 82}, 011121 (2010). 

\bibitem{marcus}
A. Chechkin and I. Pavlyukevich, J. Phys. A: Math. Theor. {\bf 47}, 342001 (2014). 

\bibitem{sche}
D. Schertzer, M. Larchev\^{e}que, J. Duan, V. V. Yanovsky, and S. Lovejoy, 
J. Math. Phys. {\bf 42}, 200 (2001). 

\end{thebibliography}
\end{document}